\documentclass[%
 reprint,
 amsmath,amssymb,
 aps,tikz
]{revtex4-1}

\usepackage[colorlinks = true, linkcolor = blue, urlcolor = blue, citecolor = blue, anchorcolor = blue]{hyperref}
\usepackage{amsmath,amsthm,amssymb,physics,color}
\usepackage{graphicx}
\usepackage{dcolumn}
\usepackage{bm}
\usepackage{lipsum}
\usepackage{soul}
\usepackage{float}

\newcommand{\iu}{{i\mkern1mu}}

\graphicspath{ {./} }

\begin{document}

\preprint{APS/123-QED}

\title{Unfolding optical transition weights of impurity materials for first-principles LCAO electronic structure calculations}

\author{Yung-Ting Lee\textsuperscript{1,2}}
 \email{ytl821@gate.sinica.edu.tw}
\author{Chi-Cheng Lee\textsuperscript{1,3}}%
\author{Masahiro Fukuda\textsuperscript{1}}%
\author{Taisuke Ozaki\textsuperscript{1}}%
\affiliation{\textsuperscript{1}Institute for Solid State Physics, The University of Tokyo, 5-1-5 Kashiwanoha, Kashiwa, Chiba 277-8581, Japan
\\
\textsuperscript{2}Institute of Atomic and Molecular Sciences, Academia Sinica, No. 1, Roosevelt Rd., Sec. 4, Taipei 10617, Taiwan
\\
\textsuperscript{3}Department of Physics, Tamkang University, No. 151, Yingzhuan Rd., Tamsui Dist., New Taipei City 251301, Taiwan
}

\date{\today}

\begin{abstract}
A method to analyze optical transitions is developed by combining the Kubo-Greenwood formula with the unfolding method to construct an unfolded electronic band structure with optical transition weights, which allows us to investigate how optical transitions are perturbed by imperfections such as impurity, vacancy, and structural distortions. Based on the Kubo-Greenwood formula, we first calculate frequency-dependent optical conductivity based on the first-principles electronic structure calculations using the linear combinations of atomic orbitals. Benefiting from the atomic orbital basis sets, the frequency-dependent optical conductivity can be traced back to their individual components before summations over all of $k$ points and bands. As a result, optical transition weights of the material can be put on the unfolded electronic band structure to show contributions at different $k$ points and bands. This method is especially useful to study the effects of broken symmetry in the optical transitions due to presence of impurities in the materials. As a demonstration, decomposed optical transition weights of a monolayer Si-doped graphene are shown in the electronic band structure.
\end{abstract}

\maketitle


\section{\label{sec:level1}Introduction}

The optical properties contain fundamental features of materials, including optical conductivity, dielectric function, refractive index, reflectivity, and transmission that can be measured by experiments \cite{exp0,exp1,expoptical1,expdielectric1,expdielectric2,exprefractiveindex0,exprefractiveindex1,exprefractiveindex2,exprefractiveindex3}, and have been widely studied for a variety of compounds, such as solids \cite{solids0,solids1,solids2,solids3}, nanoparticles \cite{nanoparticle1,nanoparticle2}, 2D materials \cite{2dmater0,2dmater1,2dmater2,2dmater3,2dmater4}, superconductors \cite{superconductor1,superconductor2,superconductor3,superconductor4}, and biological tissues \cite{biotis1}. The optical conductivity and dielectric function of materials are two important measurable quantities for understanding natural phenomena, such as current density caused by an alternating electric field, optical transitions, and energy dissipation \cite{optical0,optical1,optical2,optical4,optical_dielectric1}. To adjust light absorption capability of materials or to shift absorption energy range for designing new optical devices, fabricating different composites of materials by dopants or substitutions are possible and promising for practical applications \cite{device0,device1,device2,device3,device4}. Therefore, deeper understanding of the transitions described by the optical conductivity and dielectric function in impurity materials is obviously an important issue.

To analyze spectra of optical conductivity and dielectric function in a material, an electronic band structure is a useful analysis tool to examine whether transitions between occupied and unoccupied states occur \cite{optical1}. Since the unfolding method has been developed, an unfolded electronic band structure of impurity materials calculated by a supercell can be constructed to ease the comparison with experimental results observed by angle-resolved photoemission spectroscopy \cite{UnfoldingMethodWku,UnfoldingMethodCCLee,UnfoldingMethodTB}. However, the conventional presentation of optical conductivity of impurity materials still cannot show the direct correspondence with their band structure, although optical conductivity based on the Kubo-Greenwood formula \cite{Kubo,Greenwood} has been widely calculated by density functional theory (DFT) packages \cite{PRBOpticalCCL,Calderin,PBAllenref22,opticalpackage2,opticalpackage3,opticalpackage4}. Recently, Bianco \textit{et al.} bridged the relation between the unfolding method and the Berry curvature with Wannier functions to investigate the Berry-phase anomalous Hall conductivity of the Fe-Co alloys \cite{UnfoldBerryCurvature}. In order to build a connection between optical transitions and the electronic band structure, we propose to present optical conductivity with the unfolding method \cite{UnfoldingMethodCCLee} to put optical transition weights of a material on the unfolded electronic band structure, which is called unfolding optical transition method in the following discussions.

The enhancement of optical conductivity of silicon doped graphene (SiG) with the tunable band gap in the visible region has been proposed to improve efficiency of photovoltaic cells \cite{SiGTheo01,SiGTheo02}. In the experiments, subsequently, the graphene at a silicon-doping level of 2.7\%-4.5\% with opening a small band gap and without affecting the carrier concentration has been fabricated to enhance the performance of SiG/GaAs heterostructure solar cells in comparison with graphene/GaAs \cite{SiGExpt}. We apply the unfolding optical transition method to analyze contributions of optical transitions of SiGs in the unfolded electronic band structure for unveiling the silicon-doping effect in graphene.

This paper is organized as follows. In Sec. II, the Kubo-Greenwood formula, partial optical transitions in an electronic band structure, and unfolding optical transition are shown. In Sec. III, an example of a monolayer Si-doped graphene is given for showing changes of unfolded partial optical conductivity between Si-doped graphenes. Finally, this research work is concluded in Sec. IV.

\section{\label{sec:level2}Computational Method}

\small
In this Sec., based on the Kubo-Greenwood formula, we will discuss the formulation of (1) optical conductivity and momentum matrix element (MME), (2) partial optical conductivity, (3) unfolded partial optical conductivity, and (4) separation of unfolded partial optical conductivity. The computational order for the optical conductivity calculation is also discussed for the implementation with localized basis sets in the Sec. II A.

\normalsize
\subsection{Optical conductivity}

\small
Based on the Kubo-Greenwood formula \cite{Kubo,Greenwood}, the frequency-dependent optical conductivity tensor $\sigma_{\alpha\beta}(\omega)$ is calculated by

\footnotesize
\begin{align}
&\sigma_{\alpha\beta}(\omega) \nonumber\\
& = \frac{-\iu }{N_{k} \Omega}
\sum_{KJJ'} \frac{ f_{KJ}-f_{KJ'} }{ \epsilon_{KJ}-\epsilon_{KJ'} }
\frac{ \bra{KJ} \hat{P}_{\alpha} \ket{KJ'} \bra{KJ'} \hat{P}_{\beta} \ket{KJ} } { \epsilon_{KJ}-\epsilon_{KJ'}+\omega+\iu\eta }, \label{eq:kgf}
\end{align}

\noindent
\small
where $\hat{P_\alpha}$ is the momentum operator along $\alpha$ direction in the atomic unit, $J$ and $J'$ are indices of states, $f_{KJ}$ is the Fermi-Dirac distribution at a $k$-point $K$ and a state $J$, $\ket{KJ}$ is a Kohn-Sham eigenstate, $\epsilon$ is an eigenvalue, $\eta$ is $0^{+}$, $N_{k}$ is the total number of $k$ points, and $\Omega$ is the volume of the unit cell. When the intraband transition or the degenerate state ($\epsilon_{KJ} = \epsilon_{KJ'}$) occurs, $(f_{KJ}-f_{KJ'}) / (\epsilon_{KJ}-\epsilon_{KJ'})$ is treated as the first derivative of the occupation number with respect to the energy \cite{PBAllenref22,Calderin,UnfoldingMethodCCLee}. The MME can be evaluated by

\footnotesize
\begin{align}
&\bra{KJ} \hat{P}_{\alpha} \ket{KJ'} \nonumber
\\
& = -\iu \sum_{a} \sum_{mn} C_{m}^{KJ*} C_{n}^{KJ'} e^{-\iu\bm{K} \cdot (\bm{R}_{a}-\bm{R}_{0})} \bra{\phi_{m}(\bm{r}-\bm{R}_{a})} \nabla_{\alpha} \ket{\phi_{n}(\bm{r})}, \label{eq:mme}
\end{align}

\noindent
\small
where $\alpha$ is along $x$, $y$, or $z$ direction, $R$ is a lattice vector, $a$ is an index of cells, $m$ and $n$ are atomic orbitals' indices, and $C$ is LCAO coefficient.

Here we estimate the computational order for the calculation of optical conductivity by Eq. (\ref{eq:kgf}). The orders of operations for calculating the MME with localized basis sets and with plane wave basis sets are $O(N)$ and $O(N^{2})$, respectively, with the number of basis functions $N$ \cite{OrderNmethod,OrderNmethod2,OpenMXRef1}. After the calculation of the first MME in Eq. (\ref{eq:kgf}), the second MME can be obtained at the same time by the relation: $\bra{KJ'} \hat{P} \ket{KJ} = \bra{KJ} \hat{P} \ket{KJ'}^{*}$. Thus, the order of operations in these two MMEs with localized basis sets is $O(N)$. Furthermore, because $k$ points and two states are summation indices in Eq. (1), the orders of operations for all of $K$, $J$, and $J'$ correspond to $O(N_{k})$, $O(N)$, and $O(N)$, respectively. The total computational complexity in the frequency-dependent optical conductivity with localized basis sets is $O(N_{k}N^{3})$ in comparison with plane wave basis sets $O(N_{k}N^{4})$. Therefore, the computational effort can be reduced by utilizing localized basis sets, which is more suitable for a large-scale system.

\normalsize
\subsection{Partial optical conductivity}

\small
Since frequency-dependent optical conductivity $\sigma_{\alpha\beta}(\omega)$ is the summation over all of the $k$ points, occupied states, and unoccupied states, Eq. (\ref{eq:kgf}) can be rewritten as

\footnotesize
\begin{align}
\sigma_{\alpha\beta}(\omega) = \frac{1}{N_{k}} \sum_{KJ} \sigma_{\alpha\beta}(K,J,\omega), \label{eq:kgf2}
\end{align}

\noindent
\small
where the partial optical conductivity $\sigma_{\alpha\beta}(K,J,\omega)$ is given by

\footnotesize
\begin{align}
&\sigma_{\alpha\beta}(K,J,\omega) \nonumber\\
& \equiv \frac{-\iu }{\Omega} \sum_{J'} \frac{ f_{KJ}-f_{KJ'} }{ \epsilon_{KJ}-\epsilon_{KJ'} }
              \frac{ \bra{KJ} \hat{P_\alpha} \ket{KJ'} \bra{KJ'} \hat{P_\beta} \ket{KJ} } { \epsilon_{KJ}-\epsilon_{KJ'} + \omega+\iu\eta }. \label{eq:scpcd}
\end{align}

\noindent
\small
In Eq. (\ref{eq:scpcd}), $\omega$ is a resonance energy to excite electrons from a state $J$ to another state $J'$. The partial optical conductivity of a material along $k$ paths in the first Brillouin zone can be calculated and put on its electronic band structure in a fat band representation we will show later on.

\normalsize
\subsection{Unfolded partial optical conductivity}

\small
To analyze how the optical conductivity $\sigma_{\alpha\beta}(\omega)$ is changed by perturbations such as impurities and structural disorders, we now combine the partial optical conductivity introduced by Eq. (\ref{eq:scpcd}) with the unfolding method \cite{UnfoldingMethodCCLee}. The partial optical conductivity $\sigma_{\alpha\beta}(K,J,\omega)$ of an impurity material in a supercell can be rewritten as

\footnotesize
\begin{align}
\sigma_{\alpha\beta}(K,J,\omega) = \frac{-\iu }{\Omega} A_{KJ,KJ}^{\alpha\beta}(\omega) \label{eq:scpcd2}
\end{align}

\noindent
\small
with the spectral function tensor for the supercell defined by

\footnotesize
\begin{align}
& A_{KJ,KJ}^{\alpha\beta}(\omega) \nonumber
\\
& \equiv \bra{KJ} \sum_{J'} \frac{ f_{KJ}-f_{KJ'} }{ \epsilon_{KJ}-\epsilon_{KJ'} } \frac{ \hat{P_\alpha} \ket{KJ'} \bra{KJ'} \hat{P_\beta} } { \epsilon_{KJ}-\epsilon_{KJ'} + \omega+\iu\eta } \ket{KJ}. \label{eq:scpcd3}
\end{align}

\noindent
\small
On the other hand, the partial optical conductivity $\sigma_{\alpha\beta}(k,j,\omega)$ of a perfect crystal as a reference system has the same expression as

\footnotesize
\begin{align}
\sigma_{\alpha\beta}(k,j,\omega)
& = \frac{-\iu }{\Omega_{\text{rc}}}
\bra{kj} \sum_{j'} \frac{ f_{kj}-f_{kj'} }{ \epsilon_{kj}-\epsilon_{kj'} } \frac{ \hat{P_\alpha} \ket{kj'} \bra{kj'} \hat{P_\beta} } { \epsilon_{kj}-\epsilon_{kj'} + \omega+\iu\eta } \ket{kj} \nonumber
\\
& = \frac{-\iu }{\Omega_{\text{rc}}} A_{kj,kj}^{\alpha\beta}(\omega), \label{eq:rcpcd}
\end{align}

\noindent
\small
where $\Omega_{\text{rc}}$ is the volume of the reference cell, $\ket{kj}$ is a Kohn-Sham eigenstate at a $k$ point and a state $j$ in the reference cell, and $A_{kj,kj}^{\alpha\beta}(\omega)$ is the spectral function tensor in the reference cell. The upper-case letters in Eq. (\ref{eq:scpcd}) and the lower-case letters in Eq. (\ref{eq:rcpcd}) stand for indices in the supercell and in the reference cell, respectively.

\indent
In order to relate partial optical conductivities between the supercell and the reference cell, the unfolding method \cite{UnfoldingMethodCCLee} provides a refined approach to unfold the band structure of a supercell to the Brillouin zone of a reference cell via a spectral function. The spectral function tensor $A^{\alpha\beta}(\omega)$ is given by

\footnotesize
\begin{align}
& A^{\alpha\beta}(\omega) = \sum_{kj} A_{kj,kj}^{\alpha\beta}(\omega) = \sum_{kj} \bra{kj} \hat{A}^{\alpha\beta}(\omega) \ket{kj}. \label{eq:a1}
\end{align}

\noindent
\small
By inserting closure relations $\sum_{kmn} \ket{km} S_{mn}^{-1}(k) \bra{kn}= \hat{I}$ into $\bra{kj} \hat{A}^{\alpha\beta}(\omega) \ket{kj}$, Eq. (\ref{eq:a1}) is rewritten as

\footnotesize
\begin{align}
& \hspace*{-0.2cm} \sum_{kj} \bra{kj} \hat{A}^{\alpha\beta}(\omega) \ket{kj} \nonumber
\\
& \hspace*{-0.2cm} = \sum_{kj} \sum_{mn} \sum_{n'm'} \bra{kj}\ket{km} S_{mn}^{-1}(k) \bra{kn} \hat{A}^{\alpha\beta}(\omega) \ket{kn'} S_{n'm'}^{-1}(k) \bra{km'}\ket{kj} \nonumber
\\
& \hspace*{-0.2cm} = \sum_{k} \sum_{mn} \sum_{n'm'} S_{mn}^{-1}(k) \bra{kn} \hat{A}^{\alpha\beta}(\omega) \ket{kn'} S_{n'm'}^{-1}(k) \bra{km'}\ket{km} \label{eq:a2}
\end{align}

\noindent
\small
with the definition

\footnotesize
\begin{align}
\ket{kn} = \frac{1}{\sqrt{L}} \sum_{\textbf{R}} e^{i\textbf{k} \cdot \textbf{R}} \ket{Rn},
\end{align}

\noindent
\small
where $m$ and $n$ are indices of atomic basis functions in the reference cell, $\ket{Rn}$ is an atomic basis function in the reference cell, $L$ is the number of unit cells in the Born-von K\'{a}rm\'{a}n boundary condition, and the closure relation $\sum_{kj} \ket{kj}\bra{kj}=\hat{I}$ is required for deriving the last equation. Due to $\sum_{m'} S_{n'm'}^{-1}(k) \bra{km'}\ket{km} = \delta_{n'm}(k)$, Eq. (\ref{eq:a2}) becomes

\footnotesize
\begin{align}
\sum_{kj} \bra{kj} \hat{A}^{\alpha\beta}(\omega) \ket{kj} = \sum_{kmn} S_{mn}^{-1}(k) \bra{kn} \hat{A}^{\alpha\beta}(\omega) \ket{km}. \label{eq:a3}
\end{align}

\small
After inserting closure relations $\sum_{KJ} \ket{KJ} \bra{KJ}=\hat{I}$ for two adjacent positions of $\hat{A}^{\alpha\beta}(\omega)$ on the right-hand side of Eq. (\ref{eq:a3}), we have

\footnotesize
\begin{align}
& \sum_{kj} \bra{kj} \hat{A}^{\alpha\beta}(\omega) \ket{kj} \nonumber
\\
& = \sum_{kmn} \sum_{KJ} S_{mn}^{-1}(k) \bra{kn}\ket{KJ} \bra{KJ} \hat{A}^{\alpha\beta}(\omega) \ket{KJ} \bra{KJ}\ket{km} \nonumber
\\
& = \sum_{kmn} \sum_{KJ} S_{mn}^{-1}(k) \bra{kn}\ket{KJ} A_{KJ,KJ}^{\alpha\beta}(\omega) \bra{KJ}\ket{km}, \label{eq:rcpcd4}
\end{align}

\noindent
\small
where

\footnotesize
\begin{align}
\bra{kn}\ket{KJ} & = \sum_{N}C^{KJ}_{N} \sum_{\textbf{r}\textbf{R}} \frac{e^{-i\textbf{k} \cdot \textbf{r}}}{\sqrt{l}} \bra{rn}\ket{RN} \frac{e^{i\textbf{K} \cdot \textbf{R}}}{\sqrt{L}},
\end{align}

\noindent
\small
and 

\begin{align}
\bra{KJ}\ket{km} & = \sum_{M}C^{KJ*}_{M} \sum_{\textbf{r}'\textbf{R}'} \frac{e^{-i\textbf{K} \cdot \textbf{R}'}}{\sqrt{L}} \bra{R'M}\ket{r'm} \frac{e^{i\textbf{k} \cdot \textbf{r}'}}{\sqrt{l}}.
\end{align}

\small
For simplicity, the summations in Eq. (\ref{eq:rcpcd4}) over $k$, $j$, and $J$ are dropped. Thus, the spectral function tensor $A_{kj,kj}^{\alpha\beta}(\omega)$ in Eq. (\ref{eq:a1}) is given by

\footnotesize
\begin{align}
A_{kj,kj}^{\alpha\beta}(\omega) = \frac{L}{l} \sum_{KG} \delta_{k-G,K} W^{k}_{KJ} A_{KJ,KJ}^{\alpha\beta}(\omega) \label{eq:rcpcd2}
\end{align}

\noindent
\small
with the unfolded spectral weight

\footnotesize
\begin{align}
W^{k}_{KJ} = \sum_{MNr} e^{\iu \textbf{k}\cdot(\textbf{r}-\textbf{r}'(M))} C_{N}^{KJ} C_{M}^{KJ*} S_{0N,rm(M)}, \label{eq:rcpcd3}
\end{align}

\noindent
\small
where $L$ is the number of unit cells in a supercell, $l$ is the number of unit cells in a reference cell, $r'(M)$ and $m(M)$ refer to lattice vectors and an orbital index in the representation of the reference cell, respectively. The eigenstate $j$ in the reference cell corresponds to the unfolded eigenstate $J$ in the supercell due to $\sum_{KG} \delta_{k-G,K}$. According to Eqs. (\ref{eq:rcpcd2}) and (\ref{eq:rcpcd3}), the weight of the spectral function tensor $A_{KJ,KJ}^{\alpha\beta}(\omega)$ is determined by the phase factor $e^{i\textbf{k}\cdot(\textbf{r}-\textbf{r}'(M))}$, LCAO coefficients, and overlap matrix elements in the unfolded spectral weight $W_{KJ}^{k}$. The phase factor governs the spectral weight of  unfolding electronic band structure of a material built with a supercell. In addition, the overlap matrix elements and LCAO coefficients in a doped material may cause the reduction or enhancement of the spectral weight because the presence of impurity makes symmetry breaking. Note that this unfolded spectral weight $W^{k}_{KJ}$ collects contributions over $K$ to obtain $A_{kj,kj}^{\alpha\beta}(\omega)$ in Eq. (\ref{eq:rcpcd2}). Therefore, $A_{kj,kj}^{\alpha\beta}(\omega)$ only includes one unfolded spectral weight summed over $K$.

After calculating $A_{KJ,KJ}^{\alpha\beta}(\omega)$ in Eq. (\ref{eq:scpcd3}) and $W^{k}_{KJ}$ in Eq. (\ref{eq:rcpcd3}), the spectral function tensor $A_{kj,kj}^{\alpha\beta}(\omega)$ for the reference cell in Eq. (\ref{eq:rcpcd2}) can be evaluated. Through Eq. (\ref{eq:rcpcd2}), the band structure of the supercell is unfolded into the Brillouin zone of the reference cell with the transition weights of partial optical conductivity. Subsequently, Eqs. (\ref{eq:scpcd2}) and (\ref{eq:rcpcd2}) can be substituted into Eq. (\ref{eq:rcpcd}) to obtain an unfolded partial optical conductivity $\sigma_{\alpha\beta}(k,j,\omega)$ represented by the reference cell as follows:

\footnotesize
\begin{align}
\sigma_{\alpha\beta}(k,j,\omega) & = \frac{-\iu }{\Omega_{\text{rc}}} A_{kj,kj}^{\alpha\beta}(\omega) \nonumber
\\
& = \frac{-\iu }{\Omega_{\text{rc}}} \frac{L}{l} \sum_{KG} \delta_{k-G,K} W^{k}_{KJ} A_{KJ,KJ}^{\alpha\beta}(\omega) \nonumber
\\
& = \left ( \frac{L}{l} \right)^{2} \sum_{KG} \delta_{k-G,K} W^{k}_{KJ} \sigma_{\alpha\beta}(K,J,\omega), \label{eq:ufcd}
\end{align}

\noindent
\small
where $\Omega / \Omega_{\text{rc}}=L / l$. Finally, after summing over frequencies $\omega$ on the interval [$a$, $b$], the unfolded and integrated partial optical conductivity $\sigma_{\alpha\beta}(k,j,\omega(a:b))$ can be expressed as

\footnotesize
\begin{align}
\sigma_{\alpha\beta}(k,j,\omega(a:b)) &
\equiv \int_{a}^{b} \sigma_{\alpha\beta}(k,j,\omega) d\omega, \label{eq:ufcds}
\end{align}

\noindent
\small
The integrated unfolded partial optical conductivity gathers contributions of optical transition weights over a selected frequency range and it can be put on the unfolded electronic band structure of a material to show optical transitions at states in a fat band representation. The numerical demonstration of unfolded and integrated partial optical conductivity is provided in the Appendix.

\normalsize
\subsection{Separation of unfolded partial optical conductivity}

\small
The equation (\ref{eq:mme}) for calculating MME includes two summations over individual atomic orbitals. After rearranging the order of the summation, Eq. (\ref{eq:mme}) can be rewritten as

\footnotesize
\begin{align}
&\bra{KJ} \hat{P}_{\alpha} \ket{KJ'} = \sum_{mn} \bra{KJ} \hat{P}_{\alpha}^{mn} \ket{KJ'}, \label{eq:mme2}
\end{align}

\noindent
\small
where $m$ and $n$ are orbitals' indices and the partial MME is definded as

\footnotesize
\begin{align}
&\bra{KJ} \hat{P}_{\alpha}^{mn} \ket{KJ'} \nonumber
\\
& \equiv -\iu \sum_{a} C_{m}^{KJ*} C_{n}^{KJ'} e^{-\iu\bm{K} \cdot (\bm{R}_{a}-\bm{R}_{0})} \bra{\phi_{m}(\bm{r}-\bm{R}_{a})} \nabla_{\alpha} \ket{\phi_{n}(\bm{r})}. \label{eq:pmme}
\end{align}

\noindent
\small
By substituting Eq. (\ref{eq:mme2}) back to Eq. (\ref{eq:scpcd}), the partial optical conductivity $\sigma_{\alpha\beta}(K,J,\omega)$ can be reexpressed as

\footnotesize
\begin{align}
\sigma_{\alpha\beta}(K,J,\omega) = \sum_{mnn'm'} \sigma_{\alpha\beta}^{mnn'm'}(K,J,\omega), \label{eq:pkgf1}
\end{align}

\noindent
\small
where

\footnotesize
\begin{align}
&\sigma_{\alpha\beta}^{mnn'm'}(K,J,\omega) \nonumber
\\
& \equiv \frac{-\iu }{\Omega}
\sum_{J'} \frac{ f_{KJ}-f_{KJ'} }{ \epsilon_{KJ}-\epsilon_{KJ'} }
\frac{ \bra{KJ} \hat{P}_{\alpha}^{mn} \ket{KJ'} \bra{KJ'} \hat{P}_{\beta}^{n'm'} \ket{KJ} } { \epsilon_{KJ}-\epsilon_{KJ'}+\omega+\iu\eta }. \label{eq:pkgf2}
\end{align}

\noindent
\small
Therefore, orbital transitions of partial optical conductivity can be evaluated by assigning four individual atomic orbitals. Similarly, by using the same rearrangement for the order of the summation in MME, orbital transitions of an unfolded partial optical conductivity can be obtained by four individual atomic orbitals. The formula of the unfolded partial optical conductivity $\sigma_{\alpha\beta}(k,j,\omega)$ in Eq. (\ref{eq:ufcd}) can be rewritten as below to show the summation over all combinations of four individual atomic orbitals as follows

\footnotesize
\begin{align}
\sigma_{\alpha\beta}(k,j,\omega) = \sum_{mnn'm'} \sigma_{\alpha\beta}^{mnn'm'}(k,j,\omega),
\end{align}

\noindent
\small
where

\footnotesize
\begin{align}
\sigma_{\alpha\beta}^{mnn'm'}(k,j,\omega) = \left ( \frac{L}{l} \right)^{2} \sum_{KG} \delta_{k-G,K} W^{k}_{KJ} \sigma_{\alpha\beta}^{mnn'm'}(K,J,\omega). \label{eq:ducd}
\end{align}

\noindent
\small
According to Eq. (\ref{eq:ducd}), the individual contribution of orbital transitions of an unfolded partial optical conductivity can be separated by four assigned orbitals, such as $s$, $p$, $d$, and $f$ orbitals.

\small
\section{\label{sec:level4} Si-doped Graphene}
To demonstrate this analysis method, we provide optical conductivity of a monolayer Si-doped graphene (SiG) as an example. A monolayer SiG with a band gap and without a degradation in carrier mobility at a low doping level had been synthesized for designing optoelectronic devices \cite{SiGExpt}. The electronic band structure and optical properties of a monolayer graphene sheet with different silicon-doping levels had been reported \cite{SiGTheo01,SiGTheo1,SiGTheo2}. In this section, we demonstrate that the transition weights of optical conductivity of SiGs can be projected to corresponding electronic band structure by using the unfolding optical transition method proposed in the paper, and discuss doping effects in a supercell of graphene.

\normalsize
\subsection{Computational details}

\small
The geometry optimizations with a regular mesh of 300 Ry in real space are performed by the OpenMX code (v3.8) based on DFT \cite{OpenMXRef1,OpenMXRef2,DFT1,DFT2} with norm-conserving pseudopotentials \cite{normconservedpp} and optimized pseudo-atomic orbitals \cite{pseudoatomicorbital1} as basis sets. The optimized radial functions used are C-s2p2d1, Si-s2p2d1, and E-s2p2d2f1 for carbon, silicon, and ghost atoms, where the abbreviations of basis functions stand for (atomic symbol)-(number of radial functions for $s$, $p$, $d$, and $f$ orbitals), such as C-s2p2d1 represents each carbon atom with 2 $s$ orbitals, 2 $p$ orbitals, and 1 $d$ orbital. The cutoff radii of optimized radial functions at each C atom, Si atom, and ghost atom are 6.0, 7.0, and 13.0 bohrs, respectively. The ghost atom is included for calculating the accurate electronic band structure of conduction levels and it is placed at the center of honeycomb ring of graphene and SiGs. The exchange-correlation energy functional is treated by the generalized gradient approximation with the Perdew-Burke-Ernzerhof form \cite{GGA}. An electronic temperature of 300 K is employed to make electrons occupy eigenstates with the Fermi-Dirac function in the calculations. For all of optimizations, the force convergence criterion is $10^{-4}$ Hartree/bohr and the electronic self-consistent field criterion is $10^{-8}$ Hartree.

\begin{figure}[h!]
	\centering
	\includegraphics[scale=0.37]{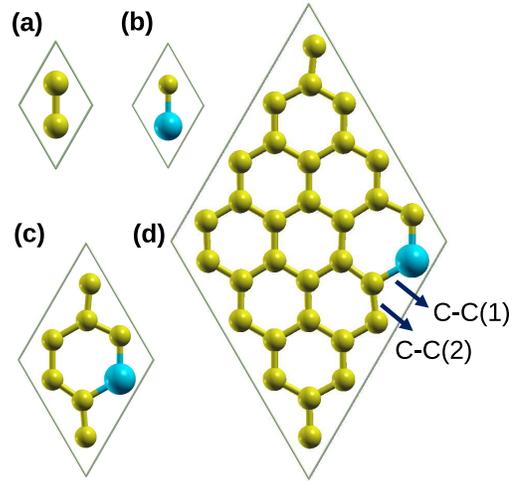}
	\caption{The top view of the optimized monolayer crystal structures by XCrySDen \cite{xcrysden}: (a) graphene, (b) SiG (1$\times$1$\times$1 supercell), (c) SiG (2$\times$2$\times$1 supercell), and (d) SiG (4$\times$4$\times$1 supercell). The yellow and cyan balls represent the C atoms and the Si atoms, respectively. Note that ghost atoms located at the center of honeycomb ring are not shown.}
	\label{fig:strfig}
\end{figure}

\begin{figure}[h!]
	\centering
	\includegraphics[scale=0.69]{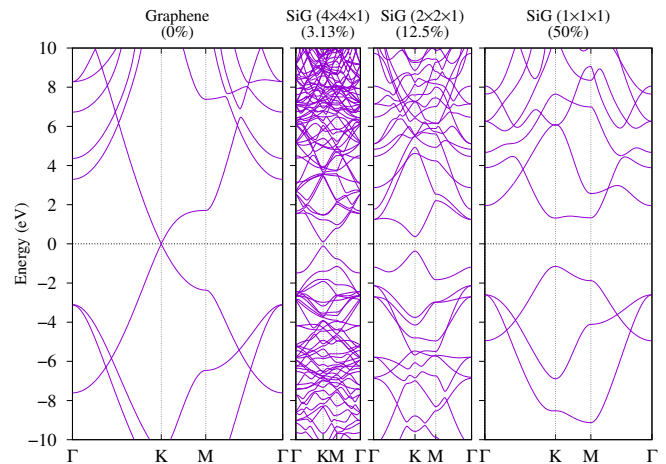}
	\caption{The electronic band structures of Si-doped graphenes (SiGs) with 0.0\%, 3.13\%, 12.5\%, and 50.0\% are shown from left to right in sequence. The Fermi level is set to be 0 eV.}
	\label{fig:bsfig}
\end{figure}

The optimized lattice constants of graphenes with different Si-doping levels are listed in Table \ref{tab:parameters} and corresponding structures are shown in Fig. \ref{fig:strfig}. By substituting a Si atom with a C atom in graphene, the structure of graphene will have a deformation due to the larger Si atomic radius \cite{atomicradius} and the longer Si-C bond length in comparison with the C-C bond lengths \cite{sicradius}. Therefore, the lattice constant $a(=b)$, Si-C, and C-C(2) bond length become longer as increasing Si concentration in graphene. These structural properties are in agreement with the experimental and calculated results \cite{SiGTheo1,graphenelattice,silicenelattice1,silicenelattice2,siglattice}. In addition, according to the electronic band structure calculations as shown in Fig. \ref{fig:bsfig}, the band gap of graphene with Si-doping of 0.00\%, 3.13\%, 12.50\%, and 50.00\% at K point of the first Brillouin zone are 0.003 eV, 0.211 eV, 0.744 eV, and 2.468 eV, respectively. As the Si-doping percentage increases, the band gap of Si-doped graphene becomes larger, which is consistent with the calculated results \cite{SiGTheo1}.

\begin{table}[h!]
\footnotesize
\caption{The optimized lattice constants, bond lengths, and $k$ meshes of graphenes with Si-doping of 0.00\%, 3.13\%, 12.50\% and 50.00\%. $a$ and $b$ refer to lattice constants at $x$-$y$ plane. The lattice constant $c$ (along $z$ axis) in the models is set to be 18 \AA. The C-C(1) and C-C(2) bond lengths (in \AA) stand for the first and second neighboring C-C bonds of the Si atom, respectively.} \label{tab:parameters}
\begin{ruledtabular}
\begin{tabular}{crccccc}
\multicolumn{2}{c}{Si-doping percentage} & \multicolumn{1}{c}{$a$(=$b$)} & C-C(1) & C-C(2) & Si-C & $k$ mesh  \\ \hline
Graphene     & 0.00 \%   & 2.467                      & 1.423   & 1.424   & -     & 24$\times$24$\times$1 \\
SiG (4$\times$4$\times$1)  & 3.13 \%   & 2.510                      & 1.409   & 1.470   & 1.681 & 6$\times$6$\times$1   \\
SiG (2$\times$2$\times$1)  & 12.50 \%  & 2.644                      & 1.437   & 1.550   & 1.692 & 12$\times$12$\times$1 \\
SiG (1$\times$1$\times$1)  & 50.00 \%  & 3.102                      & -       & -       & 1.791 & 24$\times$24$\times$1
\end{tabular}
\end{ruledtabular}
\end{table}

\normalsize
\subsection{Optical conductivity of Si-doped graphenes}

\small
Electron currents of the Si-doped graphene with a small band gap can be induced by applying a voltage to penetrate through its $x$-$y$ plane from source to drain \cite{SiGExpt}. Because of the fact that electron current density is proportional to optical conductivity, i.e. $J(\omega)=\sigma(\omega)E(\omega)$, we analyze frequency-dependent optical conductivity of SiGs to investigate the Si-doping effect by comparing it with that in non-doped graphene.

\begin{figure}[h!]
    \hspace*{-0.2cm}
	\centering
	\includegraphics[scale=0.7]{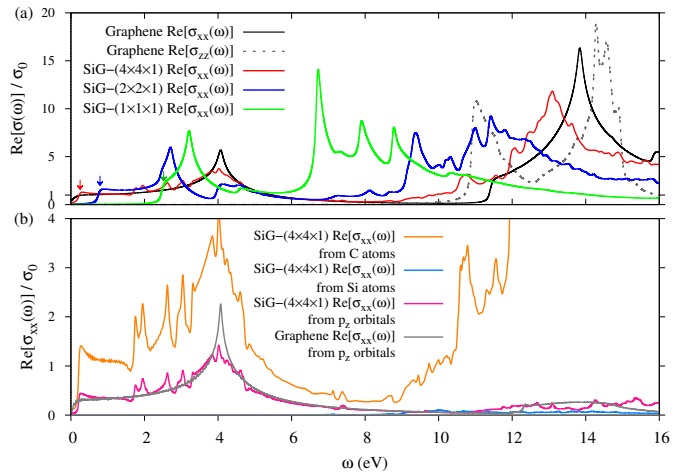}
	\caption{ The optical conductivities Re[$\sigma(\omega)$] of graphene and SiGs with doping level 3.13\%, 12.5\%, and 50.0\% are shown in (a). The partial optical conductivities Re[$\sigma_{xx}(\omega)$] of graphene and SiGs are shown in (b). The unit of conductivity $\sigma_0$ is $\text{e}^{2}/4\hbar$ \cite{opticalpackage3,sigma01}. The $k$ meshes of pristine graphene, SiG-(2$\times$2$\times$1), and SiG-(4$\times$4$\times$1) are set to be 400$\times$400$\times$1, 200$\times$200$\times$1, and 100$\times$100$\times$1, respectively, in optical calculations. The resonance energy corresponds to the energy difference between two states, such as $|E_{\text{unoccupied}} - E_{\text{occupied}}|$.}
	\label{fig:cdfig} 
\end{figure}

Using the Kubo-Greenwood formula in Eq. (\ref{eq:kgf}), the frequency-dependent optical conductivity of graphene and SiGs are calculated as shown in Fig. \ref{fig:cdfig} (a). The real part of optical conductivity $\sigma_{xx}(\omega)$(=$\sigma_{yy}(\omega)$) of graphene in $xx$/$yy$ direction is dominant at low frequencies (below 10 eV) \cite{GraExpt1,GraExpt2}, while the $zz$ component only appears above 10 eV. Optical conductivity of graphene at low frequencies is triggered by a low applied voltage. Therefore, the $zz$ component of optical conductivity of graphene materials has little contribution to electron currents and can be ignored.

In addition, the optical conductivity of graphene with a Si-doping level about 3.13\% (SiG-(4$\times$4$\times$1)) has a similar shape in comparison with that of graphene and its peak at around 4 eV is slightly weaker due to the substitution of a silicon atom for one of 32 carbon atoms in graphene in Fig. \ref{fig:cdfig} (a). As the Si-doping percentage is getting higher (more than 12.5\%), Si-doped graphenes become more like insulators gradually. The arrow-pointed peaks in Fig. \ref{fig:cdfig} (a) indicate the band gap becomes larger since the peak of optical conductivity of SiG-(4$\times$4$\times$1) shifts from 0.211 eV to 0.744 eV in SiG-(2$\times$2$\times$1) and to 2.468 eV in SiG-(1$\times$1$\times$1). It implies that optical conductivity of SiGs will decrease gradually at a low applied voltage as the Si-doping level increases.

Since SiG-(4$\times$4$\times$1) has a similar optical conductivity with graphene, we analyze individual contributions of optical conductivity decomposed to C atoms, Si atom, and/or relevant orbitals. In Fig. \ref{fig:cdfig} (b), the partial optical conductivity contributed from C atoms is almost the same as total optical conductivity of SiG-(4$\times$4$\times$1). As for the partial optical conductivity decomposed to the Si atom, the contribution of optical conductivity is quite low and close to zero at $\omega < 8$ eV, which implies that electrons within the Si atom are not induced to move on the $x$-$y$ plane. Furthermore, after optical conductivities of graphene and SiG-(4$\times$4$\times$1) were separated from all of $p_{z}$ orbitals as shown in Fig. \ref{fig:cdfig} (b) with gray and magenta lines, one can notice that the shapes of partial optical conductivity in both cases are similar although their magnitudes are lower than those decomposed to all orbitals in C atoms (with the orange line) about 30\%.

\normalsize
\subsection{Unfolded and integrated partial optical conductivity of Si-doped graphenes}

\begin{figure}[h!]
	\centering
	\includegraphics[scale=0.163]{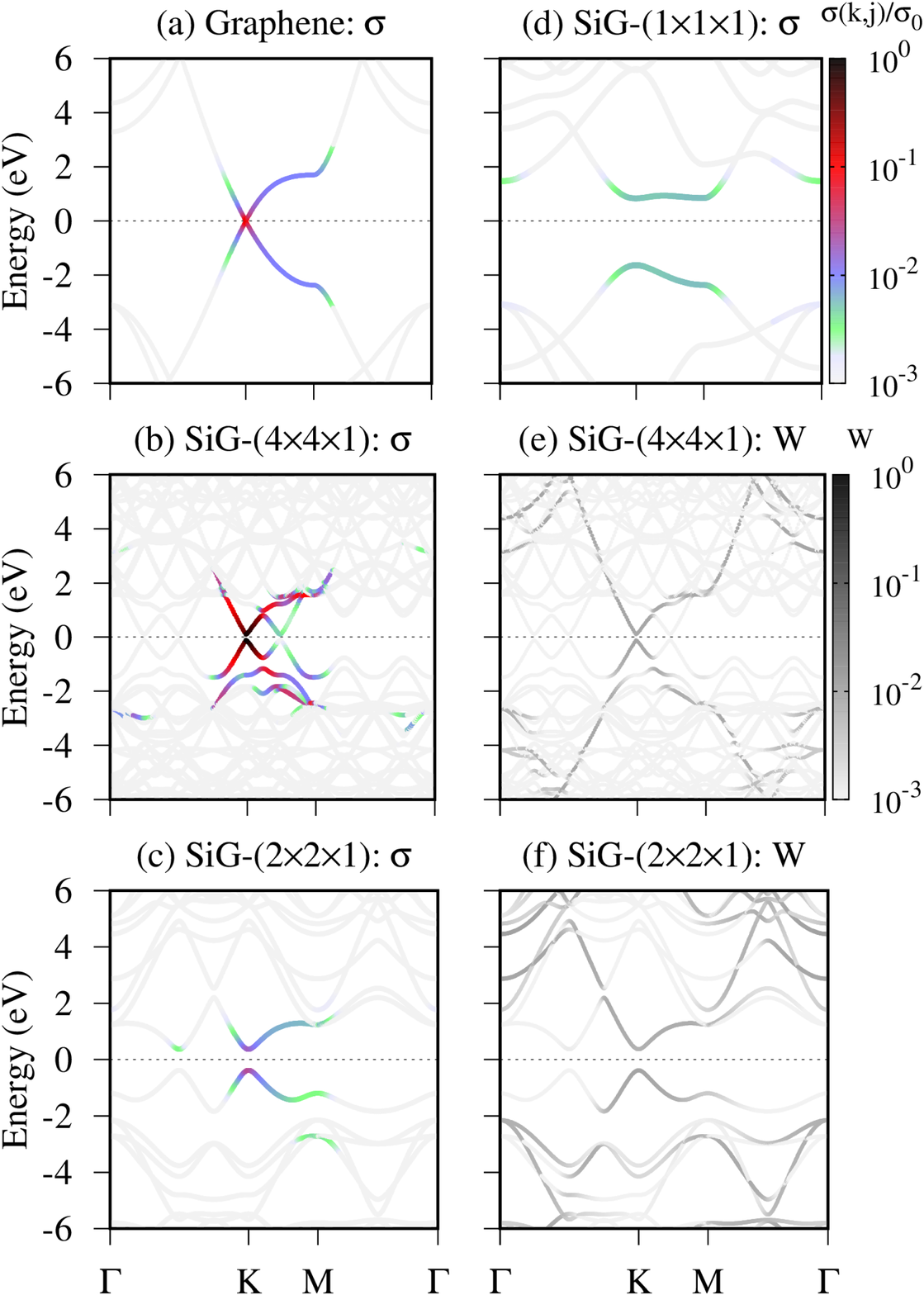}
	\caption{The real part of unfolded and integrated partial optical conductivity Re[$\sigma_{(xx+yy)/2}(k,j,\omega(0:6 \text{ eV}))$] of graphene with different Si-doping levels: (a) 0\%, (b) 3.13\% , (c) 12.5\%, and (d) 50.0\% are put in the corresponding (unfolded) band structure. The unit of unfolded and integrated partial optical conductivity is set to be the same as one in Fig. \ref{fig:cdfig}. $\eta$ is 0.05 eV. The unfolded band wegihts (W) of SiG-(4$\times$4$\times$1) and SiG-(2$\times$2$\times$1) are shown in (e) and (f), respectively.}
	\label{fig:ucdfig}
\end{figure}

Unfolded and integrated partial optical conductivity $\sigma_{\alpha\beta}(k,j,\omega(a:b))$ gives an alternative way to investigate the transition weights of optical conductivity in an impurity material at different $k$ and states after summation over frequencies $\omega$ from $a=0$ eV to $b=6$ eV by Eq. (\ref{eq:ufcds}). In order to show changes of optical transitions between graphene and SiGs, the unfolded and integrated partial optical conductivity of graphene with different Si-doping levels are calculated and shown in Fig. \ref{fig:ucdfig} (a). First, the major transition weights of partial optical conductivity of graphene in the electronic band structure at below 6 eV come from K point and M point. Graphene has a large optical transition at K point at $\omega \approx 0$ eV. Also, optical transitions between two states in graphene take place at a flat band (close to M point) and it corresponds to the sharp peak of optical conductivity in graphene at $\omega \approx$ 4 eV in Fig. \ref{fig:cdfig} (a). Second, in the case of SiG-(4$\times$4$\times$1), the optical transition occurs at K point at $\omega \approx 0.211$ eV and at the flat band ($\omega \approx 4$ eV) as shown in Fig. \ref{fig:ucdfig} (b). Graphene and SiG-(4$\times$4$\times$1) have similar pattern of optical transitions. However, due to a low Si-doping level (3.13\%), SiG-(4$\times$4$\times$1) opens a small band gap and its states from K point to M point are slightly split. Third, as the Si-doping level increases over 12.5\%, optical conductivities in SiG-(2$\times$2$\times$1) and SiG-(1$\times$1$\times$1) are getting small, although major optical transitions still occur at the K$\rightarrow$M path in Figs. \ref{fig:ucdfig} (c) and (d). It leads to decrease of the total optical conductivity of SiG at a high Si-doping percentage.

\begin{figure}[h!]
	\centering
	\includegraphics[scale=0.69]{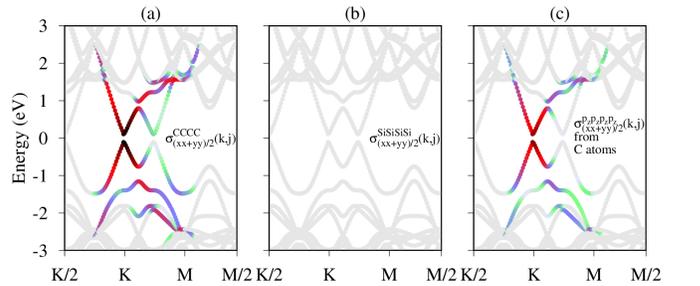}
	\caption{The real part of unfolded and integrated partial optical conductivity Re[$\sigma_{(xx+yy)/2}(k,j,\omega(0:6 \text{ eV}))$] of SiG-(4$\times$4$\times$1) decomposed to (a) the C atoms, (b) the Si atom, and (c) C atoms' $p_{z}$ orbitals are put on in the unfolded electronic band structure. The color-box scale for the unfolded and integrated partial optical conductivity is set to be the same as one in Fig. \ref{fig:ucdfig}. $\eta$ is 0.05 eV.}
	\label{fig:upcdfig}
\end{figure}

In order to investigate the Si-doping effect, the unfolded and integrated partial optical conductivity decomposed to C atoms and Si atom in the SiG-(4$\times$4$\times$1) are shown in Figs. \ref{fig:upcdfig} (a) and (b), respectively. In Fig. \ref{fig:upcdfig} (a), the transition weights of optical conductivity of SiG-(4$\times$4$\times$1) contributed from C atoms are almost the same as those of optical conductivity of SiG-(4$\times$4$\times$1) in Fig. \ref{fig:ucdfig} (b). In contrast, in Fig. \ref{fig:upcdfig} (b), the transition weights of optical conductivity of SiG-(4$\times$4$\times$1) contributed from the Si atom are quite low. It implies that optical transitions of SiG-(4$\times$4$\times$1) come from C atoms, not from the Si atom. Therefore, as the Si-doping level increases, optical conductivity of SiG will become less and its band gap will be getting larger. The Si-doping effect is like placing stones into a river to hinder current flow.

Furthermore, unfolded and integrated partial optical conductivity contributed from all of $p_{z}$ orbitals in C atoms in Fig. \ref{fig:upcdfig} (c) shows the same pattern of optical transitions as one from all orbitals in C atoms in Fig. \ref{fig:upcdfig} (a) and its contribution lowers about 30\%. In addition, the $d_{xz}$ orbitals or $d_{yz}$ orbitals also involve the $\pi$-$\pi^{*}$ transition like the transition from $p_{z}$ orbitals to $p_{z}$ orbitals, but their contributions are much lower. The magnitude order of optical transition weights belonging to orbitals in SiG-(4$\times$4$\times$1) is $W_{p_{z}-p_{z}} > W_{p_{z}-d_{xz/yz}} > W_{d_{xz}-d_{xz}} (= W_{d_{yz}-d_{yz}}) > W_{d_{xz}-d_{yz}} (= W_{d_{yz}-d_{xz}})$. Consequently, the most part of electrons can be driven by orbitals with $z$ components in C atoms to induce current flow when a low voltage is applied.

\section{\label{sec:level5} Conclusions}
We have developed an unfolding optical transition method by combining the Kubo-Greenwood formula with the unfolding method for the band structure. This unfolding optical transition method enables us to construct an unfolded electronic band structure of a supercell to a reference cell with optical transition weights, which provides an analysis tool to understand how the optical transition is perturbed by structural imperfections such as impurities and disorders. Although we developed the unfolding optical transition method for the LCAO method, it might be straightforward to apply the ideal for other methods with Wannier functions \cite{UnfoldingMethodWku,WannierFunctions,WannierFunctions2}. We have applied the method to optical conductivity of graphene with different Si-doping levels for studying the silicon-doping effect. Results show that the C atoms in the SiG-(4$\times$4$\times$1) contribute almost all of optical conductivity whereas the Si atom has little contribution after unfolded and integrated partial optical conductivity is decomposed to C atoms and the Si atom in the SiG-(4$\times$4$\times$1). It implies that doping Si atoms can decrease optical conductivity of SiGs and hinder current flow. Furthermore, after the decomposition to different orbitals by unfolding optical transition method, the $p_{z}$ orbitals of C atoms contribute the largest optical conductivity from K point to M point in the first Brillouin zone. The magnitude order of optical transition weights belonging to orbitals in the SiG-(4$\times$4$\times$1) is $W_{p_{z}-p_{z}} > W_{p_{z}-d_{xz/yz}} > W_{d_{xz}-d_{xz}} (= W_{d_{yz}-d_{yz}}) > W_{d_{xz}-d_{yz}} (= W_{d_{yz}-d_{xz}})$. These optical transitions correspond to $\pi$-$\pi^{*}$ transitions. It shows that the orbitals with $z$ components in C atoms provide main channels to make electrons flow from source to drain. Finally, in addition to the frequency-dependent optical conductivity $\sigma(\omega)$, the unfolding optical transition method provides an alternative method to present ($k$, state)-dependent optical conductivity of an impurity material in an unfolded electronic band structure for studying defects, disorders, and doping effects.

\begin{acknowledgments}
\small
This paper is partly based on results obtained from a project commissioned by the New Energy and Industrial Technology Development Organization of Japan (NEDO) Grant (P16010). Chi-Cheng Lee acknowledges partial support from the Ministry of Science and Technology of Taiwan under contract No. MOST 108-2112-M-032-010-MY2.
\end{acknowledgments}

\small
\section*{APPENDIX: Numerical demonstration of unfolded and integrated patial optical conductivity}

\begin{figure}[H]
	\includegraphics[scale=0.70]{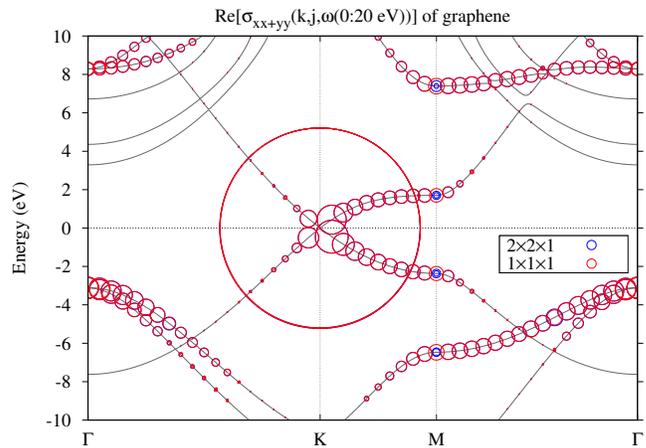}
	\caption{The (unfolded) integrated partial optical conductivities Re[$\sigma_{xx+yy}(k,j,\omega(0:20 \text{ eV}))$] of graphene (1$\times$1$\times$1) and (2$\times$2$\times$1) (with $\eta$ = 0.05 eV) are shown in the corresponding state of the electronic band structure. The transition weights of (unfolded) integrated partial optical conductivities are presented by size of blue/red circles. The solid line is the band structure of graphene.}
	\label{fig:cucd}
\end{figure}

To confirm that the unfolding optical transition method is valid, we take the unfolded and integrated partial optical conductivity Re[$\sigma_{xx+yy}(k,j,\omega(0:20 \text{ eV}))$] of Graphene-(2$\times$2$\times$1) as an example in comparison with that of Graphene-(1$\times$1$\times$1). The unfolded and integrated partial optical conductivity of graphene-(2$\times$2$\times$1) is plotted in the electronic band structure with open circles whose size is proportional to the magnitude of the Re[$\sigma_{xx+yy}(k,j,\omega(0:20 \text{ eV}))$] as shown in Fig. \ref{fig:cucd}. The transition weights of the unfolded and integrated partial optical conductivity of graphene-(2$\times$2$\times$1) are almost the same as those of graphene-(1$\times$1$\times$1), except for M point. Degenerate states with different transition weights appear, like M point, after applying the unfolding method. The sum of transition weights at these degenerate states is equal to one. The sum of unfolded and integrated partial frequency-dependent optical conductivities of graphene-(2$\times$2$\times$1) at the degenerate energy level is close to that of graphene-(1$\times$1$\times$1). Note that the small difference of the unfolded and integrated partial optical conductivity of graphene-(2$\times$2$\times$1) in comparison with the partial optical conductivity of graphene-(1$\times$1$\times$1) can be attributed to numerical error in the different unit cells.

\end{document}